\documentclass[a4paper,11pt]{article}
\usepackage{pos}
\usepackage{url}
\usepackage{amsmath}
\usepackage{cleveref}
\DeclareRobustCommand{\ion}[2]{\relax\ifmmode
                              \ifx\testbx\f at series
                              {\mathbf{#1\,\mathsc{#2}}}\else
                              {\mathrm{#1\,\mathsc{#2}}}\fi
                              \else\textup{#1\,{\mdseries\textsc{#2}}}%
                              \fi}

\newcommand{\nup}{$\nu_{\rm peak}^S$}

\crefformat{footnote}{#2\footnotemark[#1]#3}

\title{A multi-wavelength view of Active Galactic Nuclei with an emphasis on $\gamma$-rays}

\author*{Paolo Padovani}

\affiliation{European Southern Observatory, Karl-Schwarzschild-Str. 2, \\
 D-85748 Garching bei M\"unchen, Germany}
 
\emailAdd{ppadovan@eso.org}

\abstract{Active Galactic Nuclei (AGN) are remarkable astronomical sources emitting over the whole electromagnetic spectrum, 
with different bands providing unique windows on distinct sub-structures and their related physics. AGN come in a large number of 
types only partially related to intrinsic differences. I highlight here the most important AGN classes, namely jetted and non-jetted, 
radiatively efficient and inefficient, and face-on and edge-on, the source types selected by 
different bands together with the most important selection effects and biases, and the underlying emission processes, 
emphasising the $\gamma$-ray band. I then conclude with a look at some open 
issues in AGN research and at the main new astronomical facilities, which will provide us with new data to tackle them.}

\FullConference{%
  7th Heidelberg International Symposium on High-Energy Gamma-Ray Astronomy (Gamma2022)\\
  4-8 July 2022\\
  Barcelona, Spain\\}

\begin{document}
\maketitle

\section{Active Galactic Nuclei}\label{sec:AGN}

There are about 2 trillion galaxies in the Universe \citep{Conselice_2016}. Although most of them are 
believed to have a supermassive ($\gtrsim 10^6~M_{\odot}$) black hole (SMBH) at their centres, in the majority 
($\gtrsim 99 \%$) of cases the SMBH is inactive \citep{Padovani_2017}. A small minority of galaxies, however, 
become active, with their nuclei being much more powerful than the nuclei of normal galaxies. These ``active 
galactic nuclei'' therefore have an  ``additional'' component, which is now universally accepted to be due 
to the actively accreting central SMBH. Namely, matter falls onto it and in the regions close to the SMBH 
converts part of its gravitational energy into radiation. 
This leads to a number of fascinating properties, which include \citep[see][for a
  review]{Padovani_2017}: 1. very high luminosities (up to $L_{\rm bol} \approx 10^{48}$ erg s$^{-1}$),
which make AGN the most powerful non-transient sources in the Universe, hence visible up to large redshifts 
\citep[currently $z=7.642$:][]{Wang_2021}; 2. small emitting regions in most bands (down to milliparsec scales),
which imply very large energy densities; 3. strong evolution of the luminosity functions, in the sense that the 
number density and/or typical power of AGN increase with redshift, with a peak at $z \approx 2$; 4. 
broad-band emission covering the entire electromagnetic spectrum from the radio to the $\gamma$-ray band
over almost twenty orders of magnitude in frequency. 

\begin{figure}
 \center
     \includegraphics[width=0.75\textwidth]{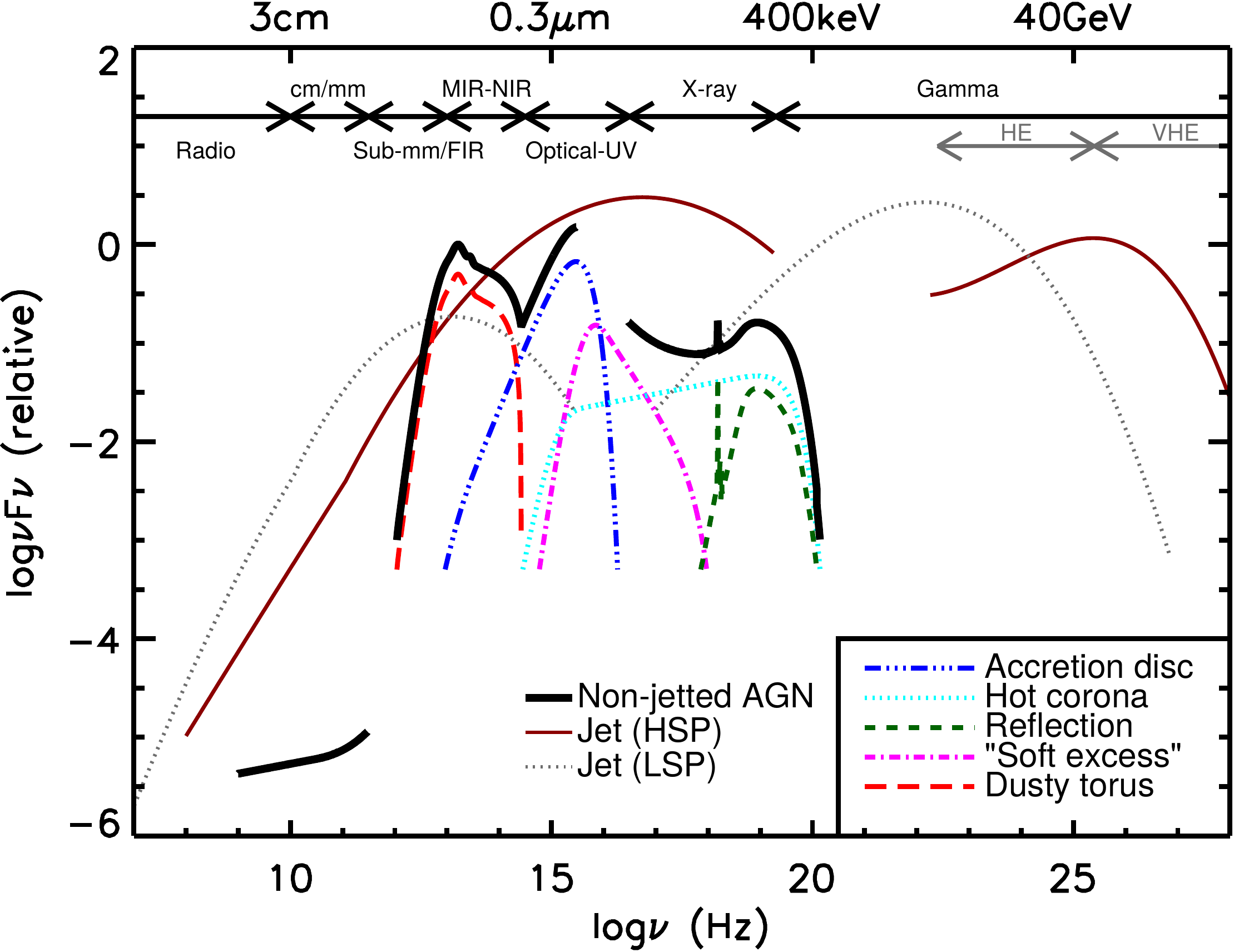}
     \caption{A schematic representation of the spectral energy distribution 
(SED) of radiatively efficient AGN, loosely based on the observed SEDs of 
non-jetted quasars 
\citep[e.g.][]{Elvis_1994,Richards_2006}. The black solid curve represents 
the total emission and the various coloured curves (shifted down for clarity) 
represent the individual components. The jet 
SED is also shown for a high-energy synchrotron peaked blazar (HSP; based on the SED of 
Mrk 421) and a low-energy synchrotron peaked blazar (LSP; based on the SED of 3C 454.3: 
see text). Reproduced from \cite{Padovani_2017}, Fig. 1, with permission, adapted 
from a figure by C. M. Harrison.}
     \label{fig:SED}
     \end{figure}

This latter property means that AGN have been, and are being, discovered 
in {\it all} spectral bands, by employing a variety of methods and 
selection techniques. The crucial point is that {\it different wavelengths 
provide different windows} on AGN physics. Namely, as discussed extensively 
in \cite{Padovani_2017}, the infrared (IR) band is mostly sensitive to obscuring 
material and dust, the optical/ultraviolet (UV) 
band is related to emission from the accretion disk (the so-called ``big blue bump''), 
while the X-ray band traces Comptonized emission from a hot corona. $\gamma$-ray and 
(high flux density) radio samples, instead, preferentially select AGN emitting strong 
non-thermal radiation coming from relativistic jets (see Fig. 
\ref{fig:SED} and Sect. \ref{sec:jetted}). 

The way AGN have been selected have led, almost unavoidably, to an explosion
of classes and sub-classes, which can be very confusing, as can be seen in 
Tab. 1 of \cite{Padovani_2017}. A small sample includes quasars, Seyfert Is and IIs, blazars, 
Fanaroff-Riley type Is and IIs, broad absorption line quasars, core-dominated quasars,
lobe-dominated quasars, Compton-tick AGN, flat and steep spectrum radio quasars, and X-ray-bright
optically normal galaxies. Reality, however, is much simpler, with most, if not all, of these 
apparently different classes being due to changes in a small number of parameters. 
Which leads me to the most relevant AGN classes.

\section{The most relevant AGN classes}
\subsection{Jetted vs. non-jetted AGN}\label{sec:jetted}

Soon after the discovery of the first quasar, 3C 273, a very strong radio source
($S_{\rm 1.4GHz} \sim 50$ Jy), it was realised that there were many more
similar sources, which were undetected by the radio telescopes of the time:
they were ``radio-quiet'' (RQ) \citep{Sandage_1965}.\footnote{Most of these
  so-called ``quasi-stellar galaxies'' turned out to be stars; but the
  concept of radio-quiet quasars, i.e., the existence of quasars with much
  weaker radio emission, proved to be correct.} These sources were later
understood to be only ``radio-faint'', as for the same optical power their
radio powers were $\approx 3$ orders of magnitude smaller than their
``radio-loud'' (RL) counterparts, but the name stuck. 

The properties of the two AGN classes in various bands have been 
compared innumerable times to try to understand their inherent differences. 
As it turned out, the distinction between the two types is 
not simply a matter of semantics but instead they represent intrinsically
different objects, with RL AGN emitting a large fraction of their
energy non-thermally and in association with strong, relativistic jets, i.e. 
structures in which matter is expelled close to the speed of light. 
The multi-wavelength emission of RQ AGN is instead dominated by thermal
emission, directly or indirectly related to the accretion disk. 
I then prefer to refer to the former as ``jetted AGN'', 
which I find much more physical than the old, but unfortunately still widespread, RL 
denomination \citep{Padovani_2017NatAs}. Note that non-jetted AGN might still have
small, weak, and slow jets, which are however of a different kind than those present
in the jetted AGN type (see also \cite{Radcliffe_2021}). 

{\it Only jetted AGN are $\gamma$-ray emitters}, which is one of the main messages 
of Fig. \ref{fig:SED} for the purpose of this symposium, which makes them particularly relevant
in this context (\cite{Padovani_2016} and Sect. \ref{sec:gamma}). Strong, relativistic jets 
are the exception, not the norm, as only (much) less than $\sim 10 \%$ of all AGN 
\citep[e.g.][]{Padovani_2011,Padovani_2015} are jetted. Jetted AGN pointing towards us 
(within, say, $10 - 15^{\circ}$) are called blazars \citep{UP_1995,Padovani_2017},
which include only roughly one galaxy out of 100,000. Despite their rarity, 
{\it blazars dominate the $\gamma$-ray sky} (Sect. \ref{sec:gamma}). 

Blazars are sub-divided from an SED point of view based on the
rest-frame frequency of their low-energy (synchrotron) hump (\nup) into
low- (LSP: \nup~$<10^{14}$~Hz), intermediate- (ISP:
$10^{14}$~Hz$ ~<$ \nup~$< 10^{15}$~Hz), and high-energy
(HSP: \nup~$> 10^{15}$~Hz) peaked sources \citep{Padovani_1995,Abdo_2010}
(see Fig. \ref{fig:SED}). 
This is quite relevant for the $\gamma$-ray band (Sect. \ref{sec:gamma}). 
Blazars come into two main flavours, i.e. flat-spectrum radio quasars (FSRQs) 
and BL Lac objects. FSRQs display broad, quasar-like emission lines 
while BL Lacs exhibit only weak, if any, absorption and emission lines
\citep{UP_1995}. Again, this is not only semantics, as these two 
classes are physically different, as detailed in the following Section. 
 
\subsection{Radiatively efficient vs. radiatively inefficient AGN}\label{sec:radiative}

The idea that not all AGN have the same central engine goes back to at least the late 70's,
when it was shown that 3CR radio galaxies could be grouped into two spectral
types, i.e. galaxies with strong [O\,{\sc ii}] 3727 and sometimes also [O\,{\sc iii}] 5007 and [Ne\,{\sc iii}] 3869 
emission lines and galaxies with only the absorption line spectra typical of giant elliptical galaxies or else very 
weak [O\,{\sc ii}] 3727 lines \citep{Hine_1979}. 

In 1980 Heckman defined ``a class of Low Ionization Nuclear Emission-line Regions (LINERs) which have
optical spectra dominated by emission-lines from low ionization species'' \citep{Heckman_1980}.  
Namely, compared to Seyferts, LINERs had stronger [O\,{\sc ii}] 3727, [O\,{\sc i}] 6300, [N\,{\sc ii}] 6584    
and weaker [O\,{\sc iii}] 5007 and [Ne\,{\sc iii}] 3869 lines. Since then other different excitation-based 
definitions have appeared in the literature but there is agreement that low-excitation AGN are less
luminous than high-excitation ones. Nowadays, the consensus is that these two classes resort to fundamentally
different ways of producing energy and a distinction is made between radiatively efficient and inefficient AGN, 
which are associated with a switch between a standard accretion, i.e. a geometrically thin (but optically thick) 
and a geometrically thick (but optically thin) accretion flow \citep[e.g.][and references therein]{Yuan_2014,Padovani_2017}.
This gets reflected into different Eddington ratios, i.e. the ratio between bolometric and the Eddington luminosity, 
$L_{\rm Edd} = 1.3 \times 10^{46}~(M/10^8 \rm M_{\odot})$ erg s$^{-1}$, 
with radiatively inefficient and efficient AGN having $L/L_{\rm Edd} \lesssim 0.01$
and $\gtrsim 0.01$ respectively \citep[e.g.][]{Heckman_2014}. In other words, the latter class 
produces more power at a given black hole mass. 
The radiatively efficient class includes all broad-lined
and narrow-lined AGN, namely quasars and Seyferts (of all orientations: Sect. \ref{sec:orientation}). 
The two blazar flavours reflect this dichotomy, with FSRQs 
being radiatively efficient and BL Lacs being inefficient (with the slight complication that ``masquerading'' BL Lacs, which are in reality FSRQs whose emission lines are washed 
out by a very bright, Doppler-boosted jet continuum, 
are therefore radiatively efficient: see, e.g., \cite{Padovani_2022} and references therein). 

Radiatively efficient AGN are the exception, which is consistent with the fact that they are more powerful than their inefficient version \citep[e.g.][]{Heckman_2014}, so inefficiency is the norm as far as AGN are concerned (not much of a surprise there ...). 

\subsection{Face-on vs. edge-on AGN}\label{sec:orientation}

Orientation plays a very big role in AGN. The seminal work by \cite{Antonucci_1985} made this very clear, 
as it showed that NGC 1068, the prototypical Seyfert II galaxy, displayed narrow lines in total light but broad lines,
like Seyfert Is, in polarised light. The authors suggested that ``the continuum source and broad line clouds are 
located inside a thick disk, with electrons above and below the disk scattering continuum and broad-line photons into the 
line of sight''. In other words, Seyfert Is and IIs are intrinsically the same objects oriented at different angles w.r.t. the line of sight.
The presence of obscuring material in a doughnut-like configuration, the so-called ``dusty torus'', 
surrounding the accretion disk on scales larger than that of the region emitting the broad lines, implies that in Seyfert IIs 
the central nucleus and close-by material are obscured by the dust and only narrow lines, emitted
by more distant clouds, are visible in their optical spectra, while in the case of Seyfert Is we have an unimpeded view of the 
accreting SMBH. This marked the birth of AGN unified schemes \citep[e.g.][]{Antonucci_1993,UP_1995}. 
According to these, broad-lined AGN (i.e., quasars and Seyfert Is) are the face-on version of narrow-lined AGN (type 2 
quasars and Seyfert IIs). 

Along the same vein, FSRQs are face-on radiatively efficient or high-excitation radio galaxies (HERGs), while 
BL Lacs are the face-on version of radiatively inefficient or low-excitation radio galaxies (LERGs). In both cases
the jet is pointing towards us (Sect. \ref{sec:jetted}) but no dust is involved in the latter. 
In fact, it now looks like dust and the broad line region (BLR) are only present at high powers ($\gtrsim 10^{42}$ erg s$^{-1}$)/high Eddington ratios
\citep[$L/L_{\rm Edd} \gtrsim 0.01$; see discussion in][]{Padovani_2017}, which implies that dust-driven unification
breaks down below these values, i.e. for radiatively inefficient AGN.

Over the last decades it has become clear that dust distribution in AGN is more complex than initially envisioned, in the sense
that multiple absorbers, on different physical scales, have to be taken into account to paint a complete view of AGN obscuration. 
Namely: 1. X-ray variability gives evidence of absorbing gas within the so-called sublimation radius, i.e. on scales smaller than the 
``dusty torus''; 2. mid-infrared interferometry has suggested a two-component structure of the torus, which includes an 
equatorial thin disk and an extended feature along the polar direction possibly due to a dusty wind; 3. increased obscuration 
with redshift might also occur on 
host galaxy scales ($\lesssim$ kpc), especially for massive ($M_* > 10^{10} M_{\odot}$) galaxies 
\citep[e.g.][and references therein]{Bianchi_2022,Honig_2017,Gilli_2022}.

Fig. 4 of \cite{Bianchi_2022} shows how the main AGN classes can be explained by 
just the three parameters discussed above: radiative efficiency, relativistic jet presence 
(or absence), and orientation.

\section{A multi-wavelength view of AGN}

I now discuss how different AGN classes are detected in the various electromagnetic bands, always with an eye to the
$\gamma$-rays.

\subsection{The radio band}\label{sec:radio}

The bright ($f_{\rm } \gtrsim 1$ mJy) GHz radio sky includes mostly jetted AGN, mainly blazars and
radio galaxies. For example, all but two of the 527 sources with $f_{\rm 5GHz} >1$ Jy and 
Galactic latitude $|b_{\rm II}| \ge 10^{\circ}$ are radio galaxies, radio quasars, or blazars, the latter
making up $\sim 51\%$ of the classified sources \citep{Kuehr_1981}. AGN selection is very easy, as one just needs 
to observe the high Galactic latitude sky. The only bias is that we are only sampling the minority
jetted AGN population. Radio emission probes the jet and is due to
synchrotron radiation (ultra-relativistic electrons moving in a magnetic field). Basically all {\it Fermi}-detected 
AGN have relatively high radio flux densities so the bright radio and $\gamma$-ray skies are very similar,
being both populated by non-thermal sources (Sect. \ref{sec:gamma}). 

When one moves to lower flux densities ($f_{\rm } \lesssim 1$ mJy) non-jetted sources become the predominant AGN type, 
as they are intrinsically fainter radio emitters, while star-forming galaxies (SFGs) become the majority population 
\citep[e.g.][and references therein]{Padovani_2016}. Once one can overcome the non-trivial separation between non-jetted AGN
and SFGs, one can reach the most common AGN with no obscuration bias (see later on). 
Radio emission at these flux density levels 
probes a variety of emission processes in AGN, i.e. those related to star-formation in the host galaxy, the corona, 
outflows, and jets, but the relative importance of these is still not clear \citep{Panessa_2019}.  

\subsection{The IR band}

As discussed in Sect. \ref{sec:orientation}, there is dust
around many AGN outside the accretion disk and on scales larger than those
of the BLR with $T \sim 100 - 1,000$
K, located at a minimum distance determined by the sublimation temperature of
the dust grains \citep[e.g.][and references therein]{Padovani_2017}. UV/optical 
emission from the accretion disk is absorbed by it and re-emitted in the infrared (IR)
band where it dominates the AGN SED at $\lambda \gtrsim 1~\mu$m up to a few tens of 
micron (Fig. \ref{fig:SED}).

AGN selected in the IR band include by definition almost only radiatively efficient AGN, 
mostly of the non-jetted type with some FSRQs, as the radiatively inefficient ones have 
no dust (Sect. \ref{sec:orientation}). The IR is sensitive to both obscured and unobscured
AGN, providing an almost isotropic selection, in particular to
extremely obscured AGN (missed by optical and soft X-ray
surveys: see below). Selection is done by typically using IR colours with the aim of
separating AGN from SFGs \citep[e.g., Sect. 3.2 in][and specifically Fig. 5
  and Table 2]{Padovani_2017}. 

\subsection{The optical/UV band}

Optical/UV emission in AGN comes from the accretion disk and the
BLR. Because of the presence of dust (Sect. \ref{sec:orientation}),
and the fact that extinction opacity is pretty large in the optical/UV band, 
this emission is detected only in unobscured
sources. The optical/UV band, therefore, provides a very biased view of the
AGN phenomenon, although it was also thanks to their strong optical/UV
emission that AGN were mostly discovered in the past.

AGN selected in the optical/UV band, therefore, include unobscured sources
mostly of the non-jetted type (as only a small fraction of jetted AGN, the 
radiatively efficient kind, also have a standard accretion disk and a BLR); in short,
broad-lined AGN. This band misses the obscured AGN (the type
2's), although many of them are still selected through their narrow
emission lines, and even the moderately obscured ones. Other biases are
against low-luminosity AGN (where the host galaxy light swamps the AGN) and
also AGN close to stellar loci (as stars are also strong optical/UV
emitters) especially at $z \sim 2.6$ and 3.5.
The optical/UV band, however, compensates for these shortcomings on two levels:
1. by providing detailed spectra, vital to study AGN physics, e.g., the
accretion disk, and the AGN spectral diversity, and to estimate the mass of the central SMBH through
``reverberation mapping''; 2. by supplying us with huge optical catalogues. 
More details on these issues can be found in Sect. 4 of \cite{Padovani_2017}.

\subsection{The X-ray band}

X-ray emission appears to be one of the defining features of AGN and
therefore the X-ray band has been crucial for AGN
studies. X-rays are supposed to be due to inverse Compton scattering of the
accretion-disk photons by an atmosphere above the disk (referred
to as the ``corona'' and whose geometry is still unknown). These X-rays
then interact with matter by being reflected, scattered, and absorbed by
the accretion disk, the dust, and even the host galaxy 
(Sect. \ref{sec:orientation}). X-ray spectra are sensitive to all of these components 
and, in particular, to the amount of absorbing material, which means they can be 
also used to classify sources into absorbed (type 2) and unabsorbed (type 1) 
\citep[e.g.][and references therein]{Bianchi_2022}. 
Low energy X-rays, in fact, are more easily absorbed than higher energy ones and so
the spectrum shape depends on the column density $N_{\rm H}$. When
$N_{\rm H} > 1.5\times10^{24}$~cm$^{-2}$ (in so-called Compton-thick [CT]
sources) the source looks completely absorbed in the X-ray band. In jetted AGN
the X-rays can have a major contribution from the jet as well. 

X-ray selected AGN, then, include in theory all AGN. However, sources with
progressively larger $N_{\rm H}$ will be systematically missed below an
increasingly higher energy, until the CT value is reached when all AGN are
undetectable in the X-ray band.  Low-luminosity AGN with $L_{\rm x}
\lesssim 10^{42}$ erg s$^{-1}$ are also biased against as this is the power
associated with host galaxy emission. LERGs are also mostly missed since their X-ray jet emission is
not very strong and they lack an accretion disk. More details on these
topics can be found in Sect. 5 of \cite{Padovani_2017}.

\subsection{The $\gamma$-ray band and the multi-messenger link}\label{sec:gamma}

Only jetted AGN, almost all of the blazar type, reach the $\gamma$-ray ($E \gtrsim 1$ 
MeV) regime, as non-jetted AGN are not detected by {\it Fermi}\footnote{
A handful of Seyferts, including NGC 1068 and NGC 4945,
are listed in the {\it Fermi-}4FGL-DR3 catalogue \citep{4LACDR3_2022} but their $\gamma$-ray 
emission is thought to be related to starburst emission in their host galaxy and not to their
SMBH \citep{Ajello_2020}.}. 
In fact, blazars make up $> 55\%$ (and likely up to $\approx 90\%$ if most unclassified sources, 
as very likely, will turn out to be blazars) of the {\it Fermi} 
(50 MeV -- 1 TeV) sky \citep[e.g.][]{4LACDR3_2022}. Moreover, $\sim 90\%$ of all extragalactic 
sources with $E > 1$ TeV are also blazars\footnote{\label{third}\url{http://tevcat.uchicago.edu/}}, studied with Imaging 
Atmospheric Cherenkov Telescopes (IACTs) such as 
MAGIC\footnote{\url{https://magic.mpp.mpg.de/}}, H.E.S.S.\footnote{\url{https://www.mpi-hd.mpg.de/hfm/HESS/}}, and 
VERITAS\footnote{\url{https://veritas.sao.arizona.edu/}}. Therefore, the $\gamma$-ray sky is similar to the bright radio
sky, as they are both dominated by blazars (Sect. \ref{sec:radio}). This is also due to the fact that blazars are relativistically
beamed, that is, Doppler boosted, which leads to an enormous increase in their observed power \citep{UP_1995}, giving them
a huge advantage over their edge-on version, namely radio galaxies. The very high-energy ($E \gtrsim 100$ GeV) 
$\gamma$-ray sky is mostly populated by a specific blazar sub-class, namely HSPs\cref{third}, as their SEDs are shifted overall
to higher frequencies (see Fig. \ref{fig:SED}).

Blazars have also recently become multi-messenger sources thanks to IceCube\footnote{\url{http://icecube.wisc.edu}}, 
the largest neutrino detector in the world, which has been operating from the South Pole for about 10 years and has 
detected hundreds of astrophysical neutrinos with energies in some cases extending beyond 1 PeV ($10^{15}$ eV). 
A high-energy neutrino event ($E \sim 290$ TeV) observed on September 22, 2017, in fact, was found to 
be in spatial coincidence with the known (``masquerading'') ISP BL Lac TXS~0506+056 at $z=0.3365$ undergoing a 
period of enhanced $\gamma$-ray emission, first observed by the {\it Fermi} satellite and 
later by the MAGIC telescopes. The chance coincidence of the neutrino alert with the flaring $\gamma$-ray source 
is at the level of $3 - 3.5\sigma$~\citep{IceCube_2018a}.
Furthermore, an archival analysis revealed that during the September 2014 to March 2015 time period
TXS 0506+056 showed a prolonged outburst when it emitted $13\pm5$ neutrinos. 
A chance correlation of this type of neutrino outburst can be excluded at a confidence 
level of $3.5\sigma$~\citep{IceCube_2018b}. This association, together with a growing body of evidence, 
links at least some blazars to IceCube neutrinos 
\citep[e.g.][and references therein]{Giommi_2020,Giommi_2021,Kurahashi_2022},
although it is also clear that only a minority of such sources can be IceCube neutrino emitters 
\citep[e.g.][]{Aartsen_2017}. This ties in with the $4.2\,\sigma$ neutrino excess recently reported 
from the direction of the local ($z = 0.004$) Seyfert 2 galaxy NGC~1068 \cite{IceCube_2022}, 
an astrophysical source, which is very different from TXS~0506+056. 

This multi-messenger connection is extremely relevant for $\gamma$-ray emitting blazars.
First, it has vital implications for the nature of $\gamma$-ray emission in blazars, which
is still debated, as the two alternative interpretations at present are not experimentally distinguishable 
\citep[e.g.][and references therein]{Cerruti_2020} (but the Cherenkov Telescope Array [CTA]
should help in this respect: e.g., \cite{Zech_2017}). In the first scenario, so-called 
leptonic, $\gamma$-ray emission is due to inverse Compton scattering between the electrons 
in the jet and their own synchrotron emission (synchrotron self-Compton) or an external photon 
field, such as the accretion disk, the broad line region, or the torus (external inverse Compton) 
\citep[e.g.][]{Maraschi_1992}. Based on the discussion above, the latter applies only to FSRQs, 
which means that within this framework the details of the dust distribution (Sect. \ref{sec:orientation}) 
will have an effect on their $\gamma$-ray emission and spectrum. In the second scenario, the hadronic 
one, $\gamma$-rays are instead thought to originate from high-energy protons either loosing energy 
through synchrotron emission or through the photo-meson process 
\citep[e.g.][]{Mannheim_1993}. The latter is the production of neutral and charged mesons ($\pi^0, \pi^+$ and $\pi^-$), 
which then decay into neutrinos, $\gamma$-rays, and other particles. Photo-meson production has one 
fundamental property: neutrinos and $\gamma$-rays (of roughly similar energy and flux) are
produced simultaneously. Neutrino detection from a blazar is the smoking gun that relativistic
protons, i.e. hadronic processes, are at work. Moreover, while $\gamma$-rays are absorbed by 
pair-production interactions with the extragalactic background light (EBL) at $E \gtrsim
100$ GeV \citep[e.g.][and references therein]{Biteau_2020}, neutrinos can travel cosmological 
distances basically unaffected by matter and magnetic fields (unlike cosmic rays) and are the only
``messengers'', which can provide information on the very high-energy physical processes
that generated them. Said differently, since the extragalactic photon sky is almost completely 
dark at the energies sampled by IceCube ($\gtrsim 60$ TeV), neutrinos are our only hope to probe
these energies. Finally, the presence of PeV neutrinos implies the existence of protons up
to energies $\gtrsim 10^{17}$\,eV. This has huge implications for the study of high-energy emission processes 
in astronomical sources. 

Note that, even compared to, e.g., IACTs, neutrino experiments still have a somewhat limited angular resolution
($\sim 0.2 - 0.3$ deg 95\% error radius in the case of NGC 1068 \cite{IceCube_2022}), making joined 
multi-messenger efforts desirable.

AGN outflows, i.e. large-scale winds of matter driven by the central SMBH, have also recently entered the $\gamma$-ray arena,
as they have been detected in sources with ultra-fast outflows (UFOs) through a stacking analysis at the $5.1 \sigma$ level 
by {\it Fermi} \cite{Ajello_2021}. These were predicted to be (faint) $\gamma$-ray sources,
as their semi-relativistic speeds (up to $\sim$ 50,000 km s$^{-1}$) can drive a shock that 
accelerates and sweeps up matter \citep[e.g.][]{King_2015}. The protons accelerated by these shocks 
can then generate low-level $\gamma$-ray emission via collisions with protons in the interstellar 
medium through proton - proton interaction followed by the meson decay described above. 

I refer to Table 3 of \cite{Padovani_2017} for a multi-wavelength overview of AGN highlighting the 
different selection biases (weaknesses) and key capabilities (strengths) of the various bands. 

\section{Open issues and the near future}\label{sec:future}

We have learnt a lot about AGN since the discovery of the first quasar in
1963. However, there are still many open questions in AGN research, 
some of them quite important, a comprehensive list of which is
given in Sect. 8.4 of \cite{Padovani_2017} (see also \cite{Antonucci_2013} for 
a critical examination of what we still do not know). The topics most relevant to this
conference include: 1. why do only a minority of AGN have jets (Sect. \ref{sec:jetted})? 
2. what is (are) the acceleration process(es) in AGN jets? 3. why are only some
blazars neutrino emitters (Sect. \ref{sec:gamma})? And what is this telling us about 
$\gamma$-ray emission processes in blazar jets? 4. what is the composition, geometry, 
and morphology of the obscuring dust (Sect. \ref{sec:orientation})? And what is its relation 
to external inverse Compton emission in FSRQs? 
 
We will soon have even more data than we had so far to tackle these and other open
issues. I list here some of the main relatively new and future facilities, sorted by band. More (still not too out-of-date) 
details and relevant hyperlinks can be found in \cite{Padovani_2017}.

\begin{itemize}

\item Radio: ASKAP (Australia), Meer\-KAT (South Africa), e-MERLIN (UK),
  WSRT - Apertif (The Netherlands), and finally the Square Kilometre Array;

\item IR: JWST (NASA/ESA), Tokyo Atacama Observatory (Japan), Euclid
  (ESA/NASA), Nancy Grace Roman Space Telescope (previously known as WFIRST; NASA);
 
\item Optical/NIR: Zwicky Transient Facility (USA), Vera C. Rubin
Observatory (previously known as LSST), and the extremely large 
  telescopes namely GMT, TMT, and the ELT;

\item X-ray: eROSITA (Germany/Russia), IXPE (NASA/ASI/+), SVOM (China/France), and eXTP (China); 

\item $\gamma$-ray: the Large High Altitude Air Shower Observatory (China) and the Cherenkov Telescope Array;
plus the missions discussed at this conference. 

\end{itemize}

Just to give a sense of how these facilities will open up entire new regions of
parameter space, especially with regard to sensitivity and number of sources, 
the Evolutionary Map of the Universe, one of the ASKAP surveys,
will detect approximately 30 million AGN in the radio band, Euclid will provide NIR spectra
for $\approx 1$ million AGN, the  Vera C. Rubin Observatory will select more than 10 million AGN, 
eROSITA will provide X-ray data for $\approx 3$ million AGN, while the
Cherenkov Telescope Array will detect $\sim 10$ times more blazars than are
currently known at TeV energies. 

In short, the future of AGN studies is very bright and we will soon be flooded with 
amazing new data. To take full advantage of them we will need to ask the right questions 
and use suitable and efficient tools.  

\section{Conclusions}

The main points of this paper can be thus summarised:

\begin{enumerate}

\item Most of the apparently different AGN classes can be explained by three features 
of their central black hole: 1. the accretion disk, which can be either radiatively efficient (the exception) 
or not (the norm) ($L/L_{\rm Edd} \gtrsim 0.01$ or $\lesssim 0.01$); 2. a strong, relativistic jet, 
which can either be there (the exception) or not (the norm); 3. orientation of the jet and the 
obscuring dust, for jetted and radiatively efficient AGN respectively.

\item Different bands give us very different perspectives on the relevant physics and distinct AGN types.
One needs to be very aware of selection effects. 

\item Jetted AGN are rare but (almost) the only $\gamma$-ray emitters; blazars rule the 
$\gamma$-ray and (the bright radio) sky because of Doppler boosting. HSPs dominate 
above $\sim 100$ GeV as a result of their SEDs. 

\item AGN (actually, blazars) have gone multi-messenger: TXS 0506+056, a blazar at z = 0.3365, 
has been associated with IceCube neutrinos. There is growing evidence that at least some blazar 
classes are neutrino sources. This is very relevant for the issue of $\gamma$-ray emission 
processes in blazars. 

\item In the next few years we will be flooded with (even more) AGN data. Hopefully they will help us to sort out a 
number of open issues in AGN research.

\end{enumerate}

{\bf Acknowledgments} I thank Paolo Giommi for useful comments on the manuscript.

\bibliographystyle{JHEP.bst}
\bibliography{my-bib-database.bib}

\providecommand{\href}[2]{#2}\begingroup\raggedright\begin{thebibliography}{10}

\bibitem{Conselice_2016}
C.J.~{Conselice}, A.~{Wilkinson}, K.~{Duncan} and A.~{Mortlock}, \emph{{The
  Evolution of Galaxy Number Density at z < 8 and Its Implications}},
  \href{https://doi.org/10.3847/0004-637X/830/2/83}{\emph{\apj} {\bfseries 830}
  (2016) 83} [\href{https://arxiv.org/abs/1607.03909}{{\ttfamily 1607.03909}}].

\bibitem{Padovani_2017}
P.~{Padovani}, D.M.~{Alexander}, R.J.~{Assef}, B.~{De Marco}, P.~{Giommi},
  R.C.~{Hickox} et~al., \emph{{Active galactic nuclei: what's in a name?}},
  \href{https://doi.org/10.1007/s00159-017-0102-9}{\emph{The Astronomy and
  Astrophysics Review} {\bfseries 25} (2017) 2}
  [\href{https://arxiv.org/abs/1707.07134}{{\ttfamily 1707.07134}}].

\bibitem{Wang_2021}
F.~{Wang}, J.~{Yang}, X.~{Fan}, J.F.~{Hennawi}, A.J.~{Barth}, E.~{Banados}
  et~al., \emph{{A Luminous Quasar at Redshift 7.642}},
  \href{https://doi.org/10.3847/2041-8213/abd8c6}{\emph{\apjl} {\bfseries 907}
  (2021) L1} [\href{https://arxiv.org/abs/2101.03179}{{\ttfamily 2101.03179}}].

\bibitem{Elvis_1994}
M.~{Elvis}, B.J.~{Wilkes}, J.C.~{McDowell}, R.F.~{Green}, J.~{Bechtold},
  S.P.~{Willner} et~al., \emph{{Atlas of Quasar Energy Distributions}},
  \href{https://doi.org/10.1086/192093}{\emph{\apjs} {\bfseries 95} (1994) 1}.

\bibitem{Richards_2006}
G.T.~{Richards}, M.~{Lacy}, L.J.~{Storrie-Lombardi}, P.B.~{Hall},
  S.C.~{Gallagher}, D.C.~{Hines} et~al., \emph{{Spectral Energy Distributions
  and Multiwavelength Selection of Type 1 Quasars}},
  \href{https://doi.org/10.1086/506525}{\emph{\apjs} {\bfseries 166} (2006)
  470} [\href{https://arxiv.org/abs/astro-ph/0601558}{{\ttfamily
  astro-ph/0601558}}].

\bibitem{Sandage_1965}
A.~{Sandage}, \emph{{The Existence of a Major New Constituent of the Universe:
  the Quasistellar Galaxies.}},
  \href{https://doi.org/10.1086/148245}{\emph{\apj} {\bfseries 141} (1965)
  1560}.

\bibitem{Padovani_2017NatAs}
P.~{Padovani}, \emph{{On the two main classes of active galactic nuclei}},
  \href{https://doi.org/10.1038/s41550-017-0194}{\emph{Nature Astronomy}
  {\bfseries 1} (2017) 0194}
  [\href{https://arxiv.org/abs/1707.08069}{{\ttfamily 1707.08069}}].

\bibitem{Radcliffe_2021}
J.F.~{Radcliffe}, P.D.~{Barthel}, M.A.~{Garrett}, R.J.~{Beswick},
  A.P.~{Thomson} and T.W.B.~{Muxlow}, \emph{{The radio emission from active
  galactic nuclei}},
  \href{https://doi.org/10.1051/0004-6361/202140791}{\emph{\aap} {\bfseries
  649} (2021) L9} [\href{https://arxiv.org/abs/2104.04519}{{\ttfamily
  2104.04519}}].

\bibitem{Padovani_2016}
P.~{Padovani}, \emph{{The faint radio sky: radio astronomy becomes
  mainstream}}, \href{https://doi.org/10.1007/s00159-016-0098-6}{\emph{The
  Astronomy and Astrophysics Review} {\bfseries 24} (2016) 13}
  [\href{https://arxiv.org/abs/1609.00499}{{\ttfamily 1609.00499}}].

\bibitem{Padovani_2011}
P.~{Padovani}, \emph{{The microjansky and nanojansky radio sky: source
  population and multiwavelength properties}},
  \href{https://doi.org/10.1111/j.1365-2966.2010.17789.x}{\emph{\mnras}
  {\bfseries 411} (2011) 1547}
  [\href{https://arxiv.org/abs/1009.6116}{{\ttfamily 1009.6116}}].

\bibitem{Padovani_2015}
P.~{Padovani}, M.~{Bonzini}, K.I.~{Kellermann}, N.~{Miller}, V.~{Mainieri} and
  P.~{Tozzi}, \emph{{Radio-faint AGN: a tale of two populations}},
  \href{https://doi.org/10.1093/mnras/stv1375}{\emph{\mnras} {\bfseries 452}
  (2015) 1263} [\href{https://arxiv.org/abs/1506.06554}{{\ttfamily
  1506.06554}}].

\bibitem{UP_1995}
C.M.~{Urry} and P.~{Padovani}, \emph{{Unified Schemes for Radio-Loud Active
  Galactic Nuclei}}, \href{https://doi.org/10.1086/133630}{\emph{\pasp}
  {\bfseries 107} (1995) 803}
  [\href{https://arxiv.org/abs/astro-ph/9506063}{{\ttfamily
  astro-ph/9506063}}].

\bibitem{Padovani_1995}
P.~{Padovani} and P.~{Giommi}, \emph{{The Connection between X-Ray-- and
  Radio-selected BL Lacertae Objects}},
  \href{https://doi.org/10.1086/175631}{\emph{\apj} {\bfseries 444} (1995) 567}
  [\href{https://arxiv.org/abs/astro-ph/9412073}{{\ttfamily
  astro-ph/9412073}}].

\bibitem{Abdo_2010}
A.A.~{Abdo}, M.~{Ackermann}, I.~{Agudo}, M.~{Ajello}, H.D.~{Aller},
  M.F.~{Aller} et~al., \emph{{The Spectral Energy Distribution of Fermi Bright
  Blazars}}, \href{https://doi.org/10.1088/0004-637X/716/1/30}{\emph{\apj}
  {\bfseries 716} (2010) 30} [\href{https://arxiv.org/abs/0912.2040}{{\ttfamily
  0912.2040}}].

\bibitem{Hine_1979}
R.G.~{Hine} and M.S.~{Longair}, \emph{{Optical spectra of 3CR radio
  galaxies.}}, \href{https://doi.org/10.1093/mnras/188.1.111}{\emph{\mnras}
  {\bfseries 188} (1979) 111}.

\bibitem{Heckman_1980}
T.M.~{Heckman}, \emph{{An Optical and Radio Survey of the Nuclei of Bright
  Galaxies - Activity in the Normal Galactic Nuclei}}, {\emph{\aap} {\bfseries
  87} (1980) 152}.

\bibitem{Yuan_2014}
F.~{Yuan} and R.~{Narayan}, \emph{{Hot Accretion Flows Around Black Holes}},
  \href{https://doi.org/10.1146/annurev-astro-082812-141003}{\emph{\araa}
  {\bfseries 52} (2014) 529} [\href{https://arxiv.org/abs/1401.0586}{{\ttfamily
  1401.0586}}].

\bibitem{Heckman_2014}
T.M.~{Heckman} and P.N.~{Best}, \emph{{The Coevolution of Galaxies and
  Supermassive Black Holes: Insights from Surveys of the Contemporary
  Universe}},
  \href{https://doi.org/10.1146/annurev-astro-081913-035722}{\emph{\araa}
  {\bfseries 52} (2014) 589} [\href{https://arxiv.org/abs/1403.4620}{{\ttfamily
  1403.4620}}].

\bibitem{Padovani_2022}
P.~{Padovani}, P.~{Giommi}, R.~{Falomo}, F.~{Oikonomou}, M.~{Petropoulou},
  T.~{Glauch} et~al., \emph{{The spectra of IceCube neutrino (SIN) candidate
  sources - II. Source characterization}},
  \href{https://doi.org/10.1093/mnras/stab3630}{\emph{\mnras} {\bfseries 510}
  (2022) 2671} [\href{https://arxiv.org/abs/2112.05394}{{\ttfamily
  2112.05394}}].

\bibitem{Antonucci_1985}
R.R.J.~{Antonucci} and J.S.~{Miller}, \emph{{Spectropolarimetry and the nature
  of NGC 1068}}, \href{https://doi.org/10.1086/163559}{\emph{\apj} {\bfseries
  297} (1985) 621}.

\bibitem{Antonucci_1993}
R.~{Antonucci}, \emph{{Unified models for active galactic nuclei and quasars}},
  \href{https://doi.org/10.1146/annurev.aa.31.090193.002353}{\emph{\araa}
  {\bfseries 31} (1993) 473}.

\bibitem{Bianchi_2022}
S.~Bianchi, V.~Mainieri and P.~Padovani, \emph{Active galactic nuclei and their
  demography through cosmic time},  in \emph{Handbook of X-ray and Gamma-ray
  Astrophysics}, C.~Bambi and A.~Santangelo, eds., (Singapore), pp.~1--32,
  Springer Nature Singapore (2022),
  \href{https://doi.org/10.1007/978-981-16-4544-0_113-1}{DOI}.

\bibitem{Honig_2017}
S.F.~{H{\"o}nig} and M.~{Kishimoto}, \emph{{Dusty Winds in Active Galactic
  Nuclei: Reconciling Observations with Models}},
  \href{https://doi.org/10.3847/2041-8213/aa6838}{\emph{\apjl} {\bfseries 838}
  (2017) L20} [\href{https://arxiv.org/abs/1703.07781}{{\ttfamily
  1703.07781}}].

\bibitem{Gilli_2022}
R.~{Gilli}, C.~{Norman}, F.~{Calura}, F.~{Vito}, R.~{Decarli}, S.~{Marchesi}
  et~al., \emph{{Supermassive black holes at high redshift are expected to be
  obscured by their massive host galaxies' interstellar medium}},
  \href{https://doi.org/10.1051/0004-6361/202243708}{\emph{\aap} {\bfseries
  666} (2022) A17} [\href{https://arxiv.org/abs/2206.03508}{{\ttfamily
  2206.03508}}].

\bibitem{Kuehr_1981}
H.~{K\"uhr}, A.~{Witzel}, I.I.K.~{Pauliny-Toth} and U.~{Nauber}, \emph{{A
  Catalogue of Extragalactic Radio Sources Having Flux Densities Greater than
  1-JY at 5-GHZ}}, {\emph{\aaps} {\bfseries 45} (1981) 367}.

\bibitem{Panessa_2019}
F.~{Panessa}, R.D.~{Baldi}, A.~{Laor}, P.~{Padovani}, E.~{Behar} and
  I.~{McHardy}, \emph{{The origin of radio emission from radio-quiet active
  galactic nuclei}},
  \href{https://doi.org/10.1038/s41550-019-0765-4}{\emph{Nature Astronomy}
  {\bfseries 3} (2019) 387} [\href{https://arxiv.org/abs/1902.05917}{{\ttfamily
  1902.05917}}].

\bibitem{4LACDR3_2022}
S.~{Abdollahi}, F.~{Acero}, L.~{Baldini}, J.~{Ballet}, D.~{Bastieri},
  R.~{Bellazzini} et~al., \emph{{Incremental Fermi Large Area Telescope Fourth
  Source Catalog}},
  \href{https://doi.org/10.3847/1538-4365/ac6751}{\emph{\apjs} {\bfseries 260}
  (2022) 53} [\href{https://arxiv.org/abs/2201.11184}{{\ttfamily 2201.11184}}].

\bibitem{Ajello_2020}
M.~{Ajello}, M.~{Di Mauro}, V.S.~{Paliya} and S.~{Garrappa}, \emph{{The
  {\ensuremath{\gamma}}-Ray Emission of Star-forming Galaxies}},
  \href{https://doi.org/10.3847/1538-4357/ab86a6}{\emph{\apj} {\bfseries 894}
  (2020) 88} [\href{https://arxiv.org/abs/2003.05493}{{\ttfamily 2003.05493}}].

\bibitem{IceCube_2018a}
{IceCube Collaboration}, M.G.~{Aartsen}, M.~{Ackermann}, J.~{Adams},
  J.A.~{Aguilar}, M.~{Ahlers} et~al., \emph{{Multimessenger observations of a
  flaring blazar coincident with high-energy neutrino IceCube-170922A}},
  \href{https://doi.org/10.1126/science.aat1378}{\emph{Science} {\bfseries 361}
  (2018) eaat1378} [\href{https://arxiv.org/abs/1807.08816}{{\ttfamily
  1807.08816}}].

\bibitem{IceCube_2018b}
{IceCube Collaboration}, M.G.~{Aartsen}, M.~{Ackermann}, J.~{Adams},
  J.A.~{Aguilar}, M.~{Ahlers} et~al., \emph{{Neutrino emission from the
  direction of the blazar TXS 0506+056 prior to the IceCube-170922A alert}},
  \href{https://doi.org/10.1126/science.aat2890}{\emph{Science} {\bfseries 361}
  (2018) 147} [\href{https://arxiv.org/abs/1807.08794}{{\ttfamily
  1807.08794}}].

\bibitem{Giommi_2020}
P.~{Giommi}, T.~{Glauch}, P.~{Padovani}, E.~{Resconi}, A.~{Turcati} and
  Y.L.~{Chang}, \emph{{Dissecting the regions around IceCube high-energy
  neutrinos: growing evidence for the blazar connection}},
  \href{https://doi.org/10.1093/mnras/staa2082}{\emph{\mnras} {\bfseries 497}
  (2020) 865} [\href{https://arxiv.org/abs/2001.09355}{{\ttfamily
  2001.09355}}].

\bibitem{Giommi_2021}
P.~{Giommi} and P.~{Padovani}, \emph{{Astrophysical Neutrinos and Blazars}},
  \href{https://doi.org/10.3390/universe7120492}{\emph{Universe} {\bfseries 7}
  (2021) 492} [\href{https://arxiv.org/abs/2112.06232}{{\ttfamily
  2112.06232}}].

\bibitem{Kurahashi_2022}
N.~{Kurahashi}, K.~{Murase} and M.~{Santander}, \emph{{High-Energy
  Extragalactic Neutrino Astrophysics}},
  \href{https://doi.org/10.1146/annurev-nucl-011122-061547}{\emph{Annual Review
  of Nuclear and Particle Science} {\bfseries 72} (2022) 365}
  [\href{https://arxiv.org/abs/2203.11936}{{\ttfamily 2203.11936}}].

\bibitem{Aartsen_2017}
M.G.~{Aartsen}, K.~{Abraham}, M.~{Ackermann}, J.~{Adams}, J.A.~{Aguilar},
  M.~{Ahlers} et~al., \emph{{The Contribution of Fermi-2LAC Blazars to Diffuse
  TeV-PeV Neutrino Flux}},
  \href{https://doi.org/10.3847/1538-4357/835/1/45}{\emph{\apj} {\bfseries 835}
  (2017) 45} [\href{https://arxiv.org/abs/1611.03874}{{\ttfamily 1611.03874}}].

\bibitem{IceCube_2022}
{IceCube Collaboration}, R.~{Abbasi}, M.~{Ackermann}, J.~{Adams},
  J.A.~{Aguilar}, M.~{Ahlers} et~al., \emph{{Evidence for neutrino emission
  from the nearby active galaxy NGC 1068}},
  \href{https://doi.org/10.1126/science.abg3395}{\emph{Science} {\bfseries 378}
  (2022) 538} [\href{https://arxiv.org/abs/2211.09972}{{\ttfamily
  2211.09972}}].

\bibitem{Cerruti_2020}
M.~{Cerruti}, \emph{{Leptonic and Hadronic Radiative Processes in
  Supermassive-Black-Hole Jets}},
  \href{https://doi.org/10.3390/galaxies8040072}{\emph{Galaxies} {\bfseries 8}
  (2020) 72} [\href{https://arxiv.org/abs/2012.13302}{{\ttfamily 2012.13302}}].

\bibitem{Zech_2017}
A.~{Zech}, M.~{Cerruti} and D.~{Mazin}, \emph{{Expected signatures from
  hadronic emission processes in the TeV spectra of BL Lacertae objects}},
  \href{https://doi.org/10.1051/0004-6361/201629997}{\emph{\aap} {\bfseries
  602} (2017) A25} [\href{https://arxiv.org/abs/1703.05937}{{\ttfamily
  1703.05937}}].

\bibitem{Maraschi_1992}
L.~{Maraschi}, G.~{Ghisellini} and A.~{Celotti}, \emph{{A Jet Model for the
  Gamma-Ray--emitting Blazar 3C 279}},
  \href{https://doi.org/10.1086/186531}{\emph{\apjl} {\bfseries 397} (1992)
  L5}.

\bibitem{Mannheim_1993}
K.~{Mannheim}, \emph{{The proton blazar.}}, {\emph{\aap} {\bfseries 269} (1993)
  67} [\href{https://arxiv.org/abs/astro-ph/9302006}{{\ttfamily
  astro-ph/9302006}}].

\bibitem{Biteau_2020}
J.~{Biteau}, E.~{Prandini}, L.~{Costamante}, M.~{Lemoine}, P.~{Padovani},
  E.~{Pueschel} et~al., \emph{{Progress in unveiling extreme particle
  acceleration in persistent astrophysical jets}},
  \href{https://doi.org/10.1038/s41550-019-0988-4}{\emph{Nature Astronomy}
  {\bfseries 4} (2020) 124} [\href{https://arxiv.org/abs/2001.09222}{{\ttfamily
  2001.09222}}].

\bibitem{Ajello_2021}
M.~{Ajello}, L.~{Baldini}, J.~{Ballet}, G.~{Barbiellini}, D.~{Bastieri},
  R.~{Bellazzini} et~al., \emph{{Gamma Rays from Fast Black-hole Winds}},
  \href{https://doi.org/10.3847/1538-4357/ac1bb2}{\emph{\apj} {\bfseries 921}
  (2021) 144} [\href{https://arxiv.org/abs/2105.11469}{{\ttfamily
  2105.11469}}].

\bibitem{King_2015}
A.~{King} and K.~{Pounds}, \emph{{Powerful Outflows and Feedback from Active
  Galactic Nuclei}},
  \href{https://doi.org/10.1146/annurev-astro-082214-122316}{\emph{\araa}
  {\bfseries 53} (2015) 115}
  [\href{https://arxiv.org/abs/1503.05206}{{\ttfamily 1503.05206}}].

\bibitem{Antonucci_2013}
R.~{Antonucci}, \emph{{Astrophysics: Quasars still defy explanation}},
  \href{https://doi.org/10.1038/495165a}{\emph{Nature} {\bfseries 495} (2013)
  165}.

\end{thebibliography}\endgroup

\end{document}